\documentclass[%
 reprint,
 amsmath,amssymb,
 aps,
]{revtex4-2}

\usepackage{graphicx}
\usepackage{dcolumn}
\usepackage{bm}

\begin{document}

\title{New Quantum Spin Perspective and Geometrical Operators of Quantum Geometry}

\author{Rakshit P. Vyas}
\email{rakshitvyas33@gmail.com} 
\author{Mihir J. Joshi}%
 
\affiliation{Department of Physics, Saurashtra University, Rajkot, India}

\date{\today}

\begin{abstract}

In this paper, we propose a new perspective of quantum spin (angular momentum) in which the Boltzmann constant \(k_{\beta}\), Planck temperature \(T_{P}\), Planck mass \(m_{P}\) and Planck area \(l_{P}^{2}\) are the integral part of the total angular momentum \(J\). With the aid of this new perspective, we modify the equation of the area and volume operator. In the quantum geometry, for \(SO(3)\) group, the angular momentum operators \(J^{k}\) is the \(k\)th Lie group generator \(T^{k}\); hence, \(T^{k} \equiv J^{k}\). Therefore, new perspective of quantum spin can be directly applicable to quantum geometry. From data, the value of the area operator \(\hat{A}_{S}\) increases with \(n^{2}\) in discrete way that suggests discrete spectrum of the area operator similar to the actual formula of the area operator. This perspective provides an \textit{auto-correct} or \textit{auto-balance} mechanism within the equation of these geometrical operators. At the quantum gravity scale, it means that the mutual small change in  \(T_{P}\), \(m_{P}\), and \(l_{P}^{2}\) occur in such a way that \(\hbar\), \(l_{P}\) and \(\hat{A}_{S}\) and \( \hat{V}_{S}\) remain invariant for a value of \(j_{i}\).  The constancy of the reduced Planck constant \(\hbar\) in the geometrical operators can provide a way through which smooth transition of the Planck scale to the nuclear or the atomic scale can be understood.


\end{abstract}

\maketitle

\section*{Introduction}

In quantum geometry or loop quantum gravity (LQG), the geometrical observable such as area and volume are quantized [1]. The spin network is very crucial to generate loop state [2,3] as well as to quantize geometrical observables [4-6] in LQG. In the spin network, Penrose [7,8] used the combinatorial principle of the angular momentum to explain space-time in discrete way. The total angular momentum is the integral part of the spin network.

Here, we give new formulation of the quantum spin using concepts of thermodynamics.  In this procedure, out of the formula of temperature, the new formula of the quantum spin is found. This new perspective has far - reaching consequences in LQG. Using this perspective, the equation of geometrical operators such as area and volume can be modified.

\section*{Quantum spin and spin network}

In thermodynamics, the temperature can be seen as average kinetic energy of considered system of particles [9] i.e., 

\begin{eqnarray}
\frac{1}{2}mv^{2} = \frac{3}{2}k_{\beta}T
\end{eqnarray}

Where \(k_{\beta}\) is the Boltzmann constant. If the equi-partition theorem is applied; then, for each degrees of freedom (\(1D\)), the kinetic energy is \(\frac{1}{2}k_{\beta}T\).  Multiplying the numerator as well as denominator of left hand side by mass \(m\) and, thereafter, multiply both sides by \(r^{2}\); one gets

\begin{eqnarray}
\frac{p^{2}r^{2}}{2m} = \frac{k_{\beta}T r^{2}}{2}
\end{eqnarray}

Since, the scalar form of angular momentum is written as \(l = rp\), the equation becomes, 

\begin{eqnarray}
\therefore l^{2} = k_{\beta}T r^{2} m
\end{eqnarray}

Till now, classical domain is considered. If Bohr's hypothesis [10] of the quantization of angular momentum is considered; then, one enters into quantum realm. i.e., 

\begin{eqnarray}
l^{2} = n^{2} \hbar^{2} = k_{\beta}T r^{2} m
\end{eqnarray}

Where \(\hbar =\frac{h}{2\pi}\), is reduced Planck constant. If \(r\), \(T\) and \(m\) are taken of the order of Planckian scale [11-13] i.e., \(r = l_{P}\), \(T = T_{P}\) and \(m = m_{P}\) respectively; one enters into the realm of the quantum gravity. Hence, 

\begin{eqnarray}
l^{2} = n^{2} \hbar^{2} = k_{\beta}T_{P} l_{P}^{2} m_{P}
\end{eqnarray}

In the spin network, the total angular momentum \(J\) is more important than the angular momentum \(l\); since, \(z\)-direction priory is unknown. In other words, the value of \(j\) plays an important role than the value of \(m\). Therefore, the total angular momentum \(J\) is used. In the spin network, the quantum spin (\(J\)) is written as twice of the actual quantum spin \(\frac{n\hbar}{2}\); i.e., \(J = 2 \times \left(\frac{n \hbar}{2}\right)\)[7,8].

Here, \(J\) is the total angular momentum. For fermions \(n\) is odd i.e., \(n = 1, 3, 5,...\) and for bosons \(n\) is even i.e., \(n = 2, 4, 6,...\)  Thus, the equation becomes,

\begin{eqnarray}
J^{2} = 2^{2} \times \left(\frac{n^{2} \hbar^{2}}{2^{2}}\right) = k_{\beta}T_{P} l_{P}^{2} m_{P}
\end{eqnarray}

By putting the value of all physical quantities in equation \(n = 1\), \(\hbar = \frac{h}{2 \pi} \rightarrow h = 6.626 \times 10^{-34} J.s\), \(k_{\beta} = 1.38 \times 10^{-23} J/K\), \(T_{P} = 1.416 \times 10^{32} K \), \(m_{P} = 2.176 \times 10^{-8} kg\) and \(l_{P} = 1.616 \times 10^{-35} m \); one can validate equality of equation. 

Hence, the value of left side of equation (6) is \(n^{2} \hbar^{2} = 1.1129 \times 10^{-68}  J^{2} \cdot s^{2} \).  The value of right side of the equation (6) is \(k_{\beta}T_{P} l_{P}^{2} m_{P} = 1.1104 \times 10^{-68}  J\cdot m^{2} \cdot kg \). Here, the numerical value and the dimension of the \(L.H.S\) and the \(R.H.S.\) of the equation (6) are same. 

Therefore, the quantum spin \(J\) in the spin network is the square root of product of Boltzmann constant \(k_{\beta}\), Planck temperature \(T_{P}\), Planck area \(l_{P}^{2}\) and Planck mass \(m_{P}\). Hence, such a quantum spin can also be called as \textit{Planck spin} at Planck scale. 

Spin network is the integral part of loop quantum gravity (LQG). Since, the spin network is the basic concept to create loop state, these results have far reaching consequences in LQG [2,3].

In equation (6), \(n = 1\) is taken; then, the reduced planck constant \(\hbar\) is, 

\begin{eqnarray}
\hbar = \sqrt{k_{\beta}T_{P} l_{P}^{2} m_{P}}
\end{eqnarray}

 Since, \(\hbar = \frac{h}{2\pi}\); the value of Planck constant is \(\vert h \vert = \sqrt{4 \pi^{2} k_{\beta}T_{P} l_{P}^{2} m_{P}} = 6.618 \times 10^{-34} J \cdot s\). Hence, this value is approximately equal to the actual value of Planck constant; that suggests an agreement with previous formalism i.e., \(h = \frac{E}{f}\).

This new perspective of quantum spin can be applied to geometrical operators of LQG such as area and volume in the following way.

\section*{The area operator of quantum geometry}

In the ADM formalism or canonical quantization of gravity, the notion of area means any \(2D\) surface embedded in any higher dimensional manifold \(\mathcal{M}\), i.e., any \(2D\) surface of \(\mathbb{R}^{2}\) embedded in \(\mathbb{R}^{3}\). Therefore,  the area of a \(2D\) surface \(S\) embedded in any manifold \(\Sigma\) is defined as [4, 6, 14, 15], 

\begin{eqnarray}
A = \int d^{2} x \sqrt{^{(2)}q}
\end{eqnarray}

Where, \(^{(2)}{q}\) is the determinant the induced metric \(^{(2)}q_{bc}\). 

The area is 2D; hence, the components of the 2D metric \(^{(2)}q_{bc}\) can be defined  in terms of the dyad basis \(e_{b}^{k}\). Here \(k,l\) are internal indices  and \(b, c \in \lbrace x, y \rbrace\) are spatial indices. 

Thus, such a 2D metric is written as [4, 6, 14, 15]

\begin{eqnarray}
^{2}q_{bc} = e_{b}^{k}e_{c}^{l}\delta_{kl}
\end{eqnarray}

If the given triad field is orthonormal, then, the transformation of equation occurs in the following way [4, 6, 14, 15],

\begin{eqnarray}
\tau_{m}^{kl}e_{z}^{m} = e_{x}^{k}e_{y}^{l}
\end{eqnarray}

For dyad basis \(\lbrace e_{b}^{k}\rbrace\), by taking the contraction of Kronecker delta  for the determinant, one gets [4, 6, 14, 15],

\begin{center}
\(det \left(^{2}q_{bc}\right) = q_{11}q_{22} - q_{12}q_{21}\)
\end{center}

\begin{center}
\( = \left( e_{x}^{k}e_{x}^{l}e_{y}^{m}e_{y}^{n} - e_{x}^{k}e_{y}^{l}e_{y}^{m}e_{x}^{n}\right) \delta_{kl}\delta_{mn}\) 
\end{center}

\begin{center}
\( = \left(\tau_{o}^{km}\tau_{p}^{ln} - \tau_{o}^{kl}\tau_{p}^{mn} \right)e_{z}^{o}e_{z}^{p}\delta_{kl}\delta_{mn} \)
\end{center}

\begin{eqnarray}
\therefore det \left(^{2}q_{bc}\right) = \vec{e}_{z}\cdot\vec{e}_{z}
\end{eqnarray}

Therefore, the classical area is written as [4, 6, 14, 15]

\begin{eqnarray}
A = \int d^{2} x \sqrt{\vec{e}_{z}\cdot\vec{e}_{z}}
\end{eqnarray}

In LQG, the frame field is conjugate momenta i.e. \(e_{z}^{k} \rightarrow - i \hbar \frac{\delta}{\delta A_{k}^{z}}\). Thus, the area operator is represented as [4, 6, 14, 15]

\begin{eqnarray}
\hat{A} = \int d^{2} x \sqrt{\delta_{kl} \frac{\delta }{\delta A_{k}^{z}}\frac{\delta }{\delta A_{l}^{z}}}
\end{eqnarray}

In LQG, the field lines are permitted to intersect or puncture; therefore, the edges of any graph \(\Theta\) intersect any surface \(S\) at many \(N\) places and point of intersection are \(\lbrace P_{1}, P_{2},...,P_{N}\rbrace\). Thus, the state \(\Psi_{\Theta}\) has area with non-zero value in the neighborhood of considered punctures. Therefore [4, 6, 14, 15],

\begin{eqnarray}
\hat{A} \Psi_ {\Theta}  \equiv \Sigma_{p = P_{1}}^{P_{N}} \sqrt{\delta_{kl} \frac{\delta }{\delta A_{k}^{z}(p)}\frac{\delta }{\delta A_{l}^{z}(p)}} \Psi_{\Theta}
\end{eqnarray}

The area operator can be applied to only at holonomy \(g_{l}\) at \(l\)th puncture. The connection \(A\) is merely relied on position \(p\). Thus, with respect to the connection, the functional derivative of the holonomy is [4, 6, 14, 15]

\begin{eqnarray}
\frac{\delta}{\delta A_{k}^{b}} g_{\lambda} [A] = n_{b}(x)T^{k} g_{\lambda} [A] = e_{b}^{k}  g_{\lambda} [A]
\end{eqnarray}

Where, \(T^{k}\) is the \(k\)th Lie group generator. At the point of the puncture  \(n^{b}\) is the unit tangent vector to the edge . For \(SO(3)\) group, \(T^{k}\) are the angular momentum operators; for instance \(T^{k} \equiv J^{k}\) (here \(J\) is angular momentum; while, \(k\) (upper indices) are internal indices)[4, 6, 14, 15].

Hence, the functional derivative takes the form [4, 6, 14, 15]

\begin{eqnarray}
\frac{\delta}{\delta A_{k}^{b}}\frac{\delta}{\delta A_{l}^{c}} \psi = n_{b}n_{c} J^{k}J^{l} \psi
\end{eqnarray}

Therefore, the equation (14) becomes [4, 6, 14, 15],

\begin{eqnarray}
\hat{A} \Psi_{\Theta} = \sum_{i} \sqrt{\delta^{kl} n_{b}n_{c} J^{k}J^{l}} \Psi_{\Theta}
\end{eqnarray}

By taking contractions over the internal and spatial indices, the equation becomes [4, 6, 14, 15], 

\begin{center}
\( \hat{A} \Psi_{\Theta} = \sum_{i} \sqrt{\delta^{kl} \hat{J}_{k}\hat{J}_{l}}\Psi_{\Theta}\) 
\end{center}

\begin{eqnarray}
\therefore \hat{A} \Psi_{\Theta} = \sum_{i} \sqrt{J^{2}} \Psi_{\Theta}
\end{eqnarray}

For rotation symmetry \(SO(3)\), \(J^{2}\) is Casimir operator. If the \(X_{b}\) are the basis of the any relevant Lie algebra, the \(X^{b}\) are the dual basis with respect to some invariant mapping of the basis; then, \(J^{2}\) is the element of the algebra \(\Sigma_{b}X_{b}X^{b}\) [4, 6, 14, 15].

For spin state,  the action of  \(J^{2}\) is [4, 6, 14, 15]

\begin{eqnarray}
J^{2} \vert j\rangle = \hbar^{2} j(j+1)\vert j\rangle
\end{eqnarray}

Since, the angular momentum is quantized; in quantum physics,  \(J^{2} = \hbar^{2} j_{i}(j_{i}+1)\) or \(J = \sqrt{j(j+1)} \hbar\).

The Casimir operator of the \(SU(2)\) group for the triad operator \(E_{k}(S)\) in the LQG is also \(T^{k}T^{k} = \hbar^{2} j(j+1)\). Hence, the quantum state for the operator \(E_{k}(S)\) is [14],

\begin{eqnarray}
E^{2}(S)\vert S\rangle = \hbar^{2}j(j+1) \vert S\rangle
\end{eqnarray}

By putting this value into equation (18), one gets,

\begin{eqnarray}
\hat{A}  = \hbar \sum_{i} \sqrt{j_{i}(j_{i}+1)} 
\end{eqnarray}

From equation (7), \(\hbar = \sqrt{k_{\beta}T_{P} l_{P}^{2} m_{P}}\); hence,

\begin{eqnarray}
\hat{A}  = \sqrt{k_{\beta}T_{P} l_{P}^{2} m_{P}} \sum_{i} \sqrt{j_{i}(j_{i}+1)} 
\end{eqnarray}

By multiplying right hand side of the equation (22) with \(8 \pi \gamma G c^{-3}\), one can restore dimensionality of both side of the equation; since, \(l_{P}^{2} = \frac{G \hbar}{c^{3}}\). Hence,

\begin{eqnarray}
\therefore \hat{A}_{S} = 8 \pi \gamma \frac{G \sqrt{k_{\beta}T_{P} l_{P}^{2} m_{P}}}{c^{3}} \sum_{i} \sqrt{j_{i}(j_{i}+1)}
\end{eqnarray}

Where, \(\gamma\) is the Barbero - Immirzi parameter, \(\hat{A}_{S}\) is the area operator for the surface \(S\). The value of \(\gamma\) is \(\gamma = 1\).

From equation (6), \(J^{2} = n^{2} \hbar^{2}\) and if If \(j_{i}(j_{i}+1)\) is defined as \(n^{2}\) i.e., \(n^{2} = j_{i}(j_{i}+1)\); then,

\begin{eqnarray}
\therefore \hat{A}_{S} = 8 \pi \gamma \frac{G \sqrt{k_{\beta}T_{P} l_{P}^{2} m_{P}}}{c^{3}} \sum \sqrt{n^{2}}
\end{eqnarray}

 The table (1) shows the value of \(n^{2}\), \(n\) and \(\hat{A}_{S}\) for the value of \(j_{i}\).

\begin{center}
\begin{tabular}{|p{0.03\textwidth} | p{0.05\textwidth} | p{0.07\textwidth}| p{0.17\textwidth} |}
\hline
\(j_{i}\) & \(n^{2}\) & \(n\) & \(\hat{A}_{S} = ...\times 10^{-70} m^{2}\) \\
\hline
\(\frac{1}{2}\) & \(\frac{3}{4}\) & \(0.8660\) & \(56.81\) \\
\hline
\(1\) & \(2\) & \(1.4142\) & \(92.77\) \\
\hline
\(\frac{3}{2}\) & \(\frac{15}{4}\) & \(1.9365\) & \(127.03\) \\
\hline
\(2\) & \(6\) & \(2.4495\) & \(160.68\) \\
\hline
\end{tabular}
\end{center}
\begin{center}
Table 1: The value of \(n^{2}\), \(n\) and \(\hat{A}_{S}\) for the value of \(j_{i}\).
\end{center}

From this data (fig. 1), the value of the area operator \(\hat{A}_{S}\) with \(n^{2}\) increases in discrete way that suggests discrete spectrum of the area operator similar to actual formula of the area operator. Since, maximum allowed value of spin in quantum physics is \(2\); here, the value of area operator is counted for \(j_{i} = \frac{1}{2}\) to \(j_{i} = 2\). From equation (7), it is found that \(T_{P}\), \(m_{P}\), and \(l_{P}^{2}\) are synchronized with each other; hence, \(\hbar\) always remains constant. This synchronization provides an \textit{auto-correct} or \textit{auto balance} mechanism that maintains \(\hbar\) constant. At quantum gravity scale, it means that mutual small change in  \(T_{P}\), \(m_{P}\), and \(l_{P}^{2}\) occur in such a way that \(\hbar\), \(l_{P}\) and \(\hat{A}_{S}\) remain invariant for a value of \(j_{i}\). Therefore, this mechanism explains invariance in the value of area operator for a value of \(j_{i}\). 
 
\begin{figure}
\begin{center}
\includegraphics[scale=0.35]{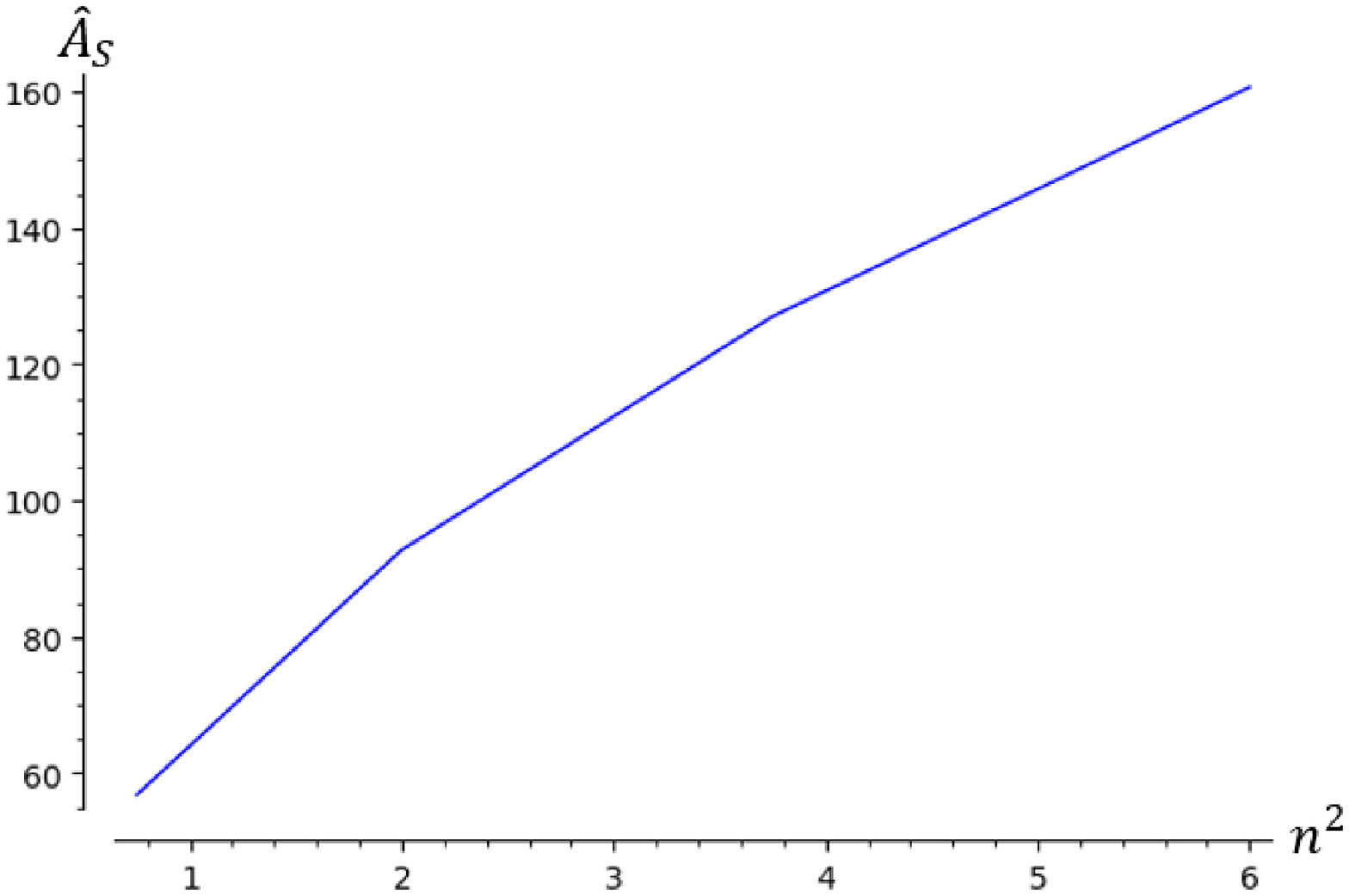}
\end{center}
\begin{center}
Fig. 1: The plot of \(\hat{A}_{S} \rightarrow n^{2}\)
\end{center}
\end{figure}

Similar to the area operator, the volume operator can also be modified in the following way.

\section*{The volume of operator of quantum geometry}

On the boundaries of volume, area of any surface are resided in the classical geometry. For any region of space i.e., \(S\), the volume can be written as [5-6, 16, 17],

\begin{eqnarray}
V = \int d^{3} x \sqrt{q}
\end{eqnarray}

Classically, it is written as [5-6, 16, 17], 

\begin{eqnarray}
V = \int d^{3} x \sqrt{ \frac{1}{3!} \left\vert \epsilon_{bcd}\epsilon^{jkl}e_{j}^{b}e_{k}^{c}e_{l}^{d} \right\vert}
\end{eqnarray}

Where, \(\epsilon_{bcd}\) is the alternating tensor.

For the volume operator, the functional derivative can be written as [5-6, 16, 17], 

\begin{eqnarray}
\frac{\delta}{\delta A_{b}^{j}}\frac{\delta}{\delta A_{c}^{k}} \frac{\delta}{\delta A_{d}^{l}} \psi = n^{b}n^{c}n^{d} J_{j}J_{k}J_{l} \psi
\end{eqnarray}

In quantum geometry, one can find the solution of Wheeler-Dewitt equation at vertices i.e., the point of intersection . Thus, one gets non-zero action of the volume operator at these vertices of the graph \(\Theta\). Hence, the given integral of the volume operator can be reduced to a sum over a finite number of vertices \(\nu \in \Theta\) that is  resided  at \(S\cap \Theta\). Thus, the volume operator becomes [5-6, 16, 17],

\begin{eqnarray}
\hat{V} \Psi_{\Theta} =  \Sigma_{\nu \in S \cup \Theta} \sqrt{\frac{1}{3!} \left\vert\epsilon_{bcd}\epsilon^{jkl}n^{b}n^{c}n^{d} \hat{J}_{j}\hat{J}_{k}\hat{J}_{l} \right\vert} \Psi_{\Theta}
\end{eqnarray}

In LQG, there are two different type of equations of volume operator i.e. (1) Rovelli-Smolin type (2) Ashtekar- Lewandowski type. The former does not include the orientation of the edges which come into the vertex; while the latter includes the orientation of the edges which come into the vertex [5-6, 16, 17]. 

The Rovelli-Smolin type of the volume operator is written as [6, 16, 17]

\begin{eqnarray}
\begin{aligned}
\hat{V}_{S}^{RS} \Psi_{\Theta} = & \ \gamma^{\frac{3}{2}} l_{P}^{3} \Sigma_{\nu \in S \cup \Theta} \Sigma_{j, k ,l} \\ & \sqrt{\left\vert\frac{i}{3!\cdot 8} \epsilon_{bcd}\epsilon^{jkl}n^{b}n^{c}n^{d} \hat{J}_{j}\hat{J}_{k}\hat{J}_{l}\right\vert}  \Psi_{\Theta}
\end{aligned}
\end{eqnarray}

Where, \(\gamma\) is Barbero-Immirzi parameter, \(l_{P}^{3}\) is Planck volume.

The Ashtekar- Lewandowski type of the volume operator is written as [5, 16, 17] 

\begin{eqnarray}
\begin{aligned}
& \ \hat{V}_{S}^{AL} \Psi_{\Theta} = \\ & \ \gamma^{\frac{3}{2}} l_{P}^{3} \Sigma_{\nu \in S \cup \Theta} \\ & \sqrt{\left\vert\frac{i}{3!\cdot 8} \epsilon_{\nu}(n^{b}, n^{c}, n^{d}) \epsilon_{bcd}\epsilon^{jkl}n^{b}n^{c}n^{d} \hat{J}_{j}\hat{J}_{k}\hat{J}_{l}\right\vert}  \Psi_{\Theta}
\end{aligned}
\end{eqnarray}

Where,  \( \epsilon_{\nu}(n^{b}, n^{c}, n^{d}) \in \lbrace -1, 0, 1 \rbrace\) is the orientation of the three tangent vectors at vertices \(v\) to the three edges meet together at \(v\). 

If orientation at vertice \(v\) is ignored and \(\hbar = \sqrt{k_{\beta}T_{P} l_{P}^{2} m_{P}}\) is taken from the equation (7); then, the general form of the volume operator is, 

\begin{eqnarray}
\begin{aligned}
\therefore \hat{V}_{S}  = & \ \gamma^{\frac{3}{2}} \left(\frac{G}{c^{3}}\right)^{\frac{3}{2}} \left(\sqrt{k_{\beta}T_{P} l_{P}^{2} m_{P}}\right)^{\frac{3}{2}} \\ & \Sigma_{\nu \in S \cup \Theta}  \sqrt{\left\vert\frac{i}{3!\cdot 8} \epsilon_{bcd}\epsilon^{jkl}n^{b}n^{c}n^{d} \hat{J}_{j}\hat{J}_{k}\hat{J}_{l}\right\vert}
\end{aligned}  
\end{eqnarray}

Similar to the area operator, the mutual  small change in  \(T_{P}\), \(m_{P}\), and \(l_{P}\) happens in such a way that \(\hbar\), \(l_{P}^{3}\) and \(\hat{V}_{S}\) remain invariant for a value of \(j_{i}\). Therefore, this mechanism also explains invariance in the value of the volume operator for a value of \(j_{i}\). 
 
\section*{The transition of the quantum geometry from the Planck scale to the nuclear scale}

This new quantum spin perspective can also explain smooth transition of the quantum geometry from the Planck scale to the nuclear scale and the atomic scale without changing the state of geometric operators such as area and volume operator.  

Furthermore, the value of \(\hbar\) for every fermion whether it is proton, neutron or electron, is constant.  For instance, the value of \(\hbar\) for proton is,

\begin{eqnarray}
\vert\hbar\vert = \sqrt{k_{\beta} T_{p^{+}} r_{p^{+}}^{2} m_{p^{+}} } =  1.054 \times 10^{-34} J \cdot s 
\end{eqnarray}

Where, \(m_{p^{+}} = 1.6726 \times 10^{-27} kg\), \(T_{p^{+}} = 0.68 \times 10^{12} K\) - (from equation (6))  and \(r_{p^{+}} = 0.84184 \times 10^{-15} m \) for proton [18].  

The equation of the area and volume operator for proton can be written by putting \(\hbar = \sqrt{k_{\beta} T_{p^{+}} r_{p^{+}}^{2} m_{p^{+}} }\) from equation (32).  Hence, the equation (23) and (31) takes the form 

\begin{eqnarray}
\hat{A}_{S} = 8 \pi \gamma \frac{G \sqrt{k_{\beta} T_{p^{+}} r_{p^{+}}^{2} m_{p^{+}}}}{c^{3}} \sum_{i} \sqrt{j_{i}(j_{i}+1)}
\end{eqnarray} 

\begin{eqnarray}
\begin{aligned}
\therefore \hat{V}_{S}  = & \ \gamma^{\frac{3}{2}} \left(\frac{G}{c^{3}}\right)^{\frac{3}{2}} \left(\sqrt{k_{\beta} T_{p^{+}} r_{p^{+}}^{2} m_{p^{+}}}\right)^{\frac{3}{2}} \\ & \Sigma_{\nu \in S \cup \Theta} \sqrt{\left\vert\frac{i}{3!\cdot 8} \epsilon_{bcd}\epsilon^{jkl}n^{b}n^{c}n^{d} \hat{J}_{j}\hat{J}_{k}\hat{J}_{l}\right\vert}
\end{aligned}  
\end{eqnarray}

For proton, the value of the quantum spin is \(j_{i}=\frac{1}{2}\); hence, the value of the area and volume operator can be calculated from equation (32) and (33). Here, the value of temperature \(\left(T_{P} \rightarrow T_{p^{+}} \right)\), mass \(\left(m_{P} \rightarrow m_{p^{+}} \right)\), and square of length \(\left(l_{P}^{2} \rightarrow r_{p^{+}}^{2}\right)\) within root changes in the equation (31); but, the reduced Planck constant \(\hbar\) remains constant. This leaves the Planck area \(l_{P}^{2}\) and the Planck volume \(l_{P}^{3}\) unchanged. Hence, the value of the area operator and the volume operator remains constant for a given spin value \(j_{i} = \frac{1}{2}\). Due to this mechanism, the quantum geometry makes smooth transition from the Planck scale to the nuclear scale without changing the state of the area and volume operator. Similar to proton, the value of area and volume operator of other fermions can also be calculated.  

\section*{Conclusions}

Finally, we conclude that, this new perspective of the quantum spin has far reaching consequences in LQG. The quantum spin \(J\) in the spin network is the square root of product of Boltzmann constant \(k_{\beta}\), Planck temperature \(T_{P}\), Planck area \(l_{P}^{2}\) and Planck mass \(m_{P}\). With the aid of this new perspective of quantum spin, the actual equation of area and volume operator can be modified. For \(SO(3)\) group, the angular momentum operators \(J^{k}\) is the \(k\)th Lie group generator \(T^{k}\); hence, \(T^{k} \equiv J^{k}\).  From \(\hbar =\sqrt{k_{\beta}T_{P} l_{P}^{2} m_{P}}\), it is found that \(T_{P}\), \(m_{P}\), and \(l_{P}^{2}\) are synchronized with each other; hence, \(\hbar\) always remains constant. This synchronization provides an \textit{auto-correct}  or \textit{auto balance} mechanism that maintains \(\hbar\) constant. At the quantum gravity scale, it means that the mutual small change in  \(T_{P}\), \(m_{P}\), and \(l_{P}^{2}\) occur in such a way that \(\hbar\), \(l_{P}\) and \(\hat{A}_{S}\) and \(\hat{V}_{S}\) remain invariant for a value of \(j_{i}\). Furthermore, the value of quantized area \(\hat{A}_{S}\) with \(n^{2}\) increases in discrete way that suggests discrete spectrum of the area operator similar to actual formula of the area operator. The smooth transition of the quantum geometry from the Planck scale to the nuclear scale (and atomic scale) can be explained without changing the state of geometric operators such as area and volume operator using this new quantum spin perspective.  Hence, the quantum geometry provides background on which the other mass particles move or interact. Hence, the constancy of the reduced Planck constant \(\hbar\) in the geometrical operators such as area and volume operator through this perspective can provide a way through which smooth transition of the Planck scale to the nuclear or the atomic scale can be understood.  
  
\section*{Acknowledgement}

The authors are thankful to Physics Department, Saurashtra University, Rajkot, India.


\begin{thebibliography}{18}

\bibitem {Rovelli}
C. Rovelli, “Loop Quantum Gravity”, \textit{Living Rev. Relativity} \textbf{11}, 5 (2008).

\bibitem{Rovelli}
C. Rovelli and L. Smolin, "Loop space representation of quantum general relativity", \textit{Nucl. Phys. B.} \textbf{B331}, pp. 80 -152, (1990).

\bibitem{Rovelli}
C. Rovelli and L. Smolin, "Spin networks and quantum gravity", \textit{Phys. Rev. D} \textbf{52}(10), (1995).

\bibitem{Ashtekar}
A. Ashtekar and J. Lewandowski, “Quantum Theory of Gravity I: Area Operators”, arXiv:gr-qc/9602046 (1996).

\bibitem{Ashtekar}
A. Ashtekar, and J. Lewandowski, “Quantum Theory of Geometry II: Volume Operators”, arXiv:gr-qc/9711031, (1997). 

\bibitem{Rovelli}
C. Rovelli and  L. Smolin “Discreteness of Area and Volume in Quantum Gravity”, arXiv:gr-qc/9411005, (1994).

\bibitem{Penrose}
R. Penrose, \textit{"On the Nature of Quantum Geometry"}, Magic Without Magic, Freeman, San Francisco, pp. 333-354, (1972).

\bibitem{Penrose}
R. Penrose, \textit{"Angular momentum: An approach to combinatorial space-time"}, Quantum Theory and Beyond, Cambridge University Press, pp. 151-180, (1971).

\bibitem{Greiner}
W. Greiner \(et\) \(al.\), \textit{Thermodynamics and Statistical Mechanics}, (Springer - Verlag, New York, U.S.A. 1995).

\bibitem{Bohr}
N. Bohr, "On the constitution of atoms and molecules", \textit{Philos. Mag.} \textbf{26}(6), 1–25, (1913).

\bibitem{Planck}
M. Planck, "Über irreversible Strahlungsvorgänge", Schöpf, HG. (eds) Von Kirchhoff bis Planck, (1978).

\bibitem{Tomilin}
K. Tomilin, \textit{"Natural Systems of Units. To the Centenary Anniversary of the Planck System"}, Proceedings Of The XXII Workshop On High Energy Physics And Field Theory,  pp. 287–296, (1999).

\bibitem{Adler}
R. Adler, "Six easy roads to the Planck scale", \textit{Am. J. Phys.} \textbf{78}(9), (2010).

\bibitem{Doná}
P. Doná, and S. Speziale, Introductory Lectures to loop Quantum Gravity”, arXiv:1007.0402 (2010); C. Rovelli, \textit{Quantum Gravity}, (Cambridge University Press, New York, 2004).

\bibitem{Ashtekar}
A. Ashtekar and J. Lewandowski, “Differential Geometry on the Space of Connections via Graphs and Projective Limits”, arXiv:hep-th/9412073, (1996).

\bibitem{Bianchi}
E. Bianchi, P. Dona’ and S. Speziale, “Polyhedra in Loop Quantum gravity”, arXiv:1009.3402, (2010). 

\bibitem{ Haggard}
H. Haggard, “Pentahedral Volume, Chaos, and Quantum Gravity”, arXiv:1211.7311, (2012).

\bibitem{ Castelvecchi}
D. Castelvecchi, "proton Size Puzzle leaps closer to resolution", \textit{Nature}, \textbf{575}, 269-270, (2019).

\end{thebibliography}
\end{document}